\documentclass[%
 table,
 superscriptaddress,
 amsmath,amssymb,
 aps,
 pra,
 twocolumn,
floatfix,
]{revtex4-2}

\usepackage[utf8]{inputenc}

\usepackage{hyperref}

\usepackage{natbib}
\usepackage[dvipsnames]{xcolor}
\usepackage{tikz}
\usetikzlibrary{automata, arrows}
\usetikzlibrary{decorations.pathreplacing,angles,quotes} 
\usetikzlibrary{positioning}
\usetikzlibrary{shapes,arrows}
\usetikzlibrary{decorations.shapes}
\usetikzlibrary{arrows.meta}
\usetikzlibrary{backgrounds, fit}
\usetikzlibrary{calc}
\tikzstyle{decision} = [diamond, draw, fill=blue!20, 
text width=4.5em, text badly centered, node distance=3cm, inner sep=0pt]
\tikzstyle{block} = [rectangle, draw, fill=blue!20, 
text width=5em, text centered, rounded corners, minimum height=4em]
\tikzstyle{line} = [draw, -latex']
\tikzstyle{cloud} = [draw, ellipse,fill=red!20, node distance=3cm,
minimum height=2em]

\usepackage{graphicx}
\usepackage[position=top, caption=false]{subfig}
\usepackage{enumitem}
\usepackage{algpseudocode}
\usepackage{algorithm}
\usepackage{amssymb}
\usepackage{mathtools}
\usepackage{bbold}
\usepackage{booktabs}
\usepackage{amsthm}
\usepackage{amsmath}
\usepackage{float}

\usepackage[hang,flushmargin]{footmisc}

\theoremstyle{definition}

\newcommand{\ket}[1]{\ensuremath{\left| #1 \right\rangle}}

\begin{document}
\title{Generation and verification of 27-qubit Greenberger-Horne-Zeilinger states in a superconducting quantum computer
}
\affiliation{School of Physics, University of Melbourne, VIC, Parkville, 3010, Australia.}
\affiliation{School of Mathematics and Statistics, University of Melbourne, VIC, Parkville, 3010, Australia.}

\author{Gary J. Mooney}
\email{mooneyg@unimelb.edu.au}
\affiliation{School of Physics, University of Melbourne, VIC, Parkville, 3010, Australia.}
\author{Gregory A. L. White}
\email{white.g@unimelb.edu.au}
\affiliation{School of Physics, University of Melbourne, VIC, Parkville, 3010, Australia.}
\author{Charles D. Hill}
\email{cdhill@unimelb.edu.au}
\affiliation{School of Physics, University of Melbourne, VIC, Parkville, 3010, Australia.}
\affiliation{School of Mathematics and Statistics, University of Melbourne, VIC, Parkville, 3010, Australia.}
\author{Lloyd C. L. Hollenberg}
\email{lloydch@unimelb.edu.au}
\affiliation{School of Physics, University of Melbourne, VIC, Parkville, 3010, Australia.}
\date{\today}

\begin{abstract} \label{sec:abstract}
Generating and detecting genuine multipartite entanglement (GME) of sizeable quantum states prepared on physical devices is an important benchmark for highlighting the progress of near-term quantum computers. A common approach to certify GME is to prepare a Greenberger-Horne-Zeilinger (GHZ) state and measure a GHZ fidelity of at least 0.5. We measure the fidelities using multiple quantum coherences of GHZ states on 11 to 27 qubits prepared on the IBM Quantum \emph{ibmq\_montreal} device. Combinations of quantum readout error mitigation (QREM) and parity verification error detection are applied to the states. A fidelity of $0.546 \pm 0.017$ was recorded for a 27-qubit GHZ state when QREM was used, demonstrating GME across the full device with a confidence level of 98.6\%. We benchmarked the effect of parity verification on GHZ fidelity for two GHZ state preparation embeddings on the heavy-hexagon architecture. The results show that the effect of parity verification, while relatively modest, led to a detectable improvement of GHZ fidelity.
\end{abstract}
\maketitle
\section*{Introduction} \label{sec:introduction}
In the race to scale-up quantum computers and demonstrate quantum advantage, an important technical milestone is the generation of entanglement across a device. Entanglement -- or the inability to factorise a multi-qubit system into separable states -- is typically seen as the essence of what differentiates quantum behaviour from classical. Indeed, it has been shown that quantum systems with low or no amounts of entanglement can be simulated efficiently on a classical computer \cite{vidal2003efficient, verstraete2004renormalization, verstraete2008matrix}. For this reason, the ability to generate and maintain genuine multipartite entanglement (GME) is fundamental for quantum information processors (QIPs) to outperform classical computers. \par 
There are various benchmarks that indicate the capabilities of a given QIP. For example, quantum volume \cite{cross2019validating} is a holistic number that takes into account qubit number and error rates of a device. Although bipartite entanglement can be rigorously quantified, it remains an open problem to do the same for GME. To demonstrate and validate GME, we select a state that is known to be entangled across multiple qubits and assess how well a device can construct that state. 
In particular, Greenberger-Horne-Zeilinger (GHZ) states are well suited to this purpose -- in that they are GME states whose fidelity on a QIP can be efficiently estimated. The fidelity estimate comes from a combination of measuring the populations of the qubit states as well as their coherences. An approach used to detect GME is to use GME witness operators~\cite{toth2005entanglement,lewenstein2000optimization, terhal2000bell}. A negative expectation value with respect to a target state is a sufficient but not necessary condition for the state containing GME. It has been shown that measuring a GHZ fidelity of at least 0.5 is equivalent to measuring a negative GME witness expectation value, hence implying that the state exhibits GME~\cite{leibfried2005creation}.\par  

\begin{figure*}
	\centering
    \includegraphics[width=0.55\linewidth]{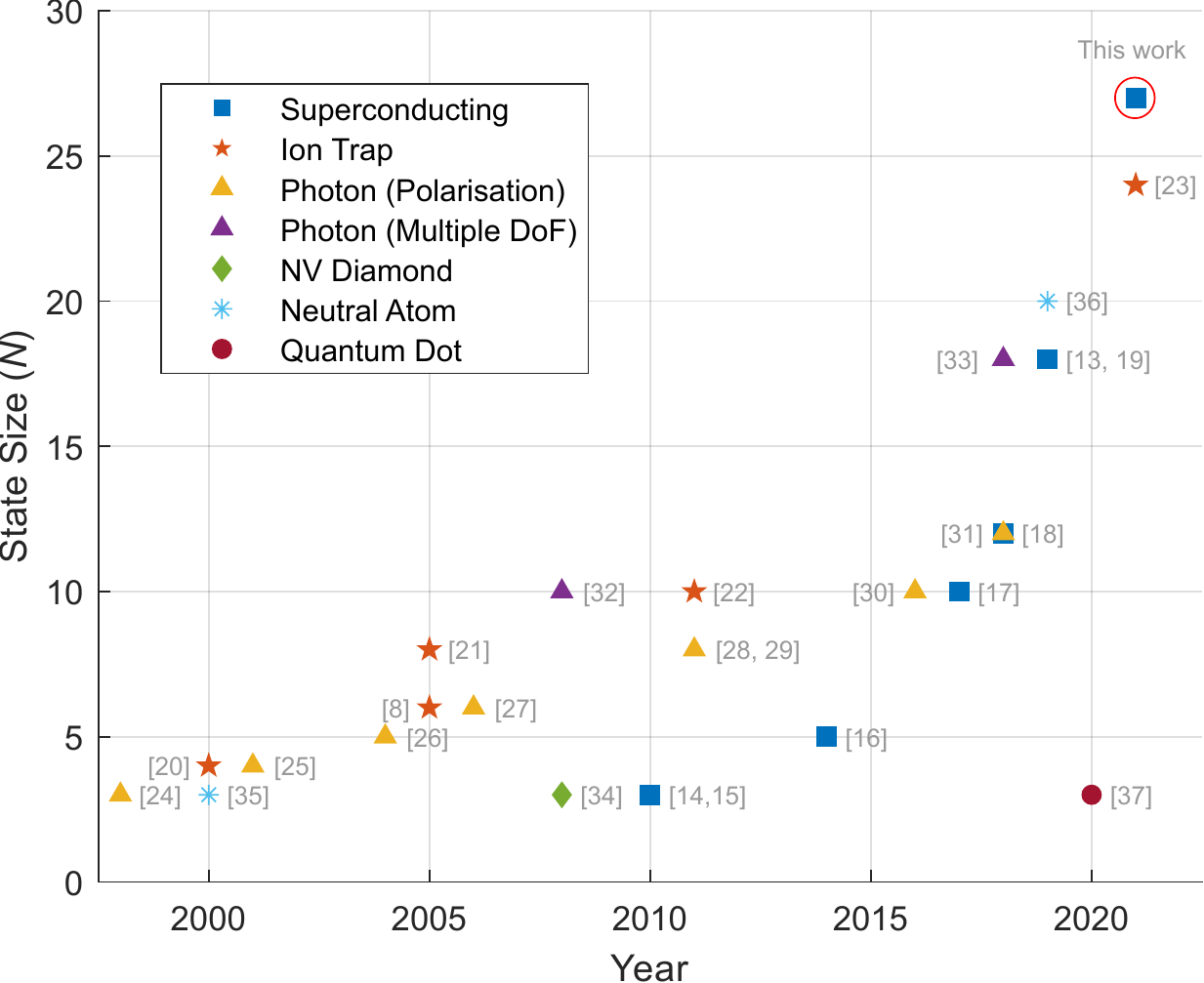}
	\caption{History of experimentally prepared quantum states exhibiting $N$-qubit GME, where $N \geq 3$, with at least 95\% confidence in gate-based quantum systems. The year is the date of first publication, to the best of our knowledge. The plot includes superconducting~\cite{neeley2010generation,dicarlo2010preparation,barends2014superconducting,song201710,gong2019genuine,wei2020verifying,song2019generation}, ion trap~\cite{sackett2000experimental,leibfried2005creation,haffner2005scalable,monz201114,pogorelov2021compact}, photonic (polarisation)~\cite{bouwmeester1999observation, pan2001experimental,zhao2004experimental,lu2007experimental,yao2012observation,huang2011experimental,wang2016experimental,zhong201812}, photonic (multiple degrees of freedom (DoF))~\cite{gao2010experimental,wang201818}, nitrogen-vacancy (NV) centres in diamond~\cite{neumann2008multipartite}, neutral atom~\cite{rauschenbeutel2000step,omran2019generation}, and quantum dot~\cite{takeda2020quantum} systems. The circled marker for $N=27$ in 2021 refers to the results of this work.} \label{fig:gme-history}
\end{figure*}

In this work we create large GHZ states up to 27 qubits on a physical device and measure their fidelities. The experiments are performed on the IBM Quantum \emph{ibmq\_montreal} device, which consists of 27 superconducting transmon qubits~\cite{koch2007charge}. The device is from the series of IBM Quantum Falcon processors and was recently benchmarked at having a quantum volume of 64~\cite{jurcevic2020demonstration}. When constructing states on QIPs, the actual entanglement of the states within the devices may be acceptably high, however the observed entanglement could be betrayed by erroneous measurement results. Using a quantum detector tomography (QDT) technique, measurement errors can be sufficiently understood and classically inverted to estimate the pre-measurement states~\cite{lundeen2009tomography}. We employ a well-known quantum readout error mitigation (QREM) technique to implement measurement correction within our experiments~\cite{maciejewski2020mitigation} which has previously been used in the certification of an 18-qubit GHZ state~\cite{wei2020verifying}. We investigate and justify the assumptions of this method when applied to the prepared noisy GHZ states to ensure that the corrections made do not inflate the actual fidelity of the states within the device. GME has been demonstrated copiously in the literature across many different QIP architectures. A plot summarising this history of results for state sizes $N \geq 3$ qubits within gate-based quantum systems is shown in Figure~\ref{fig:gme-history}. With QREM applied, we record a fidelity of~$0.546 \pm 0.017$ with~98.6\% confidence for being above the 0.5 threshold for a 27-qubit GHZ state, which appears to be the largest demonstration of GME to-date~\cite{neeley2010generation,dicarlo2010preparation,barends2014superconducting,song201710,gong2019genuine,wei2020verifying,song2019generation,sackett2000experimental,leibfried2005creation,haffner2005scalable,monz201114,pogorelov2021compact,bouwmeester1999observation, pan2001experimental,zhao2004experimental,lu2007experimental,yao2012observation,huang2011experimental,wang2016experimental,zhong201812,gao2010experimental,wang201818,neumann2008multipartite,rauschenbeutel2000step,omran2019generation,takeda2020quantum,wang201816,friis2018observation,mooney2019entanglement,pu2018experimental}. Error bars represent the standard error (of the mean). Beyond implementation of QREM, we investigate the use of parity verification on the fidelity of the GHZ states created. Parity verification is a fundamental error detection protocol used within quantum error correction schemes towards the realisation of large-scale fault-tolerant quantum computing. Ancilla qubits are used to measure the parity of state qubits to detect errors, enabling erroneous computations to be corrected or discarded. We benchmark the effects of parity verification on entanglement generation for various sized GHZ states prepared on the \emph{ibmq\_montreal} device. This work highlights the technical achievement in quantum hardware and the positive progress towards the realisation of practical quantum computers.

\section*{Results}

\subsection*{Detecting genuine multipartite entanglement in GHZ states}
\begin{figure}
	\centering
     \includegraphics[scale=1]{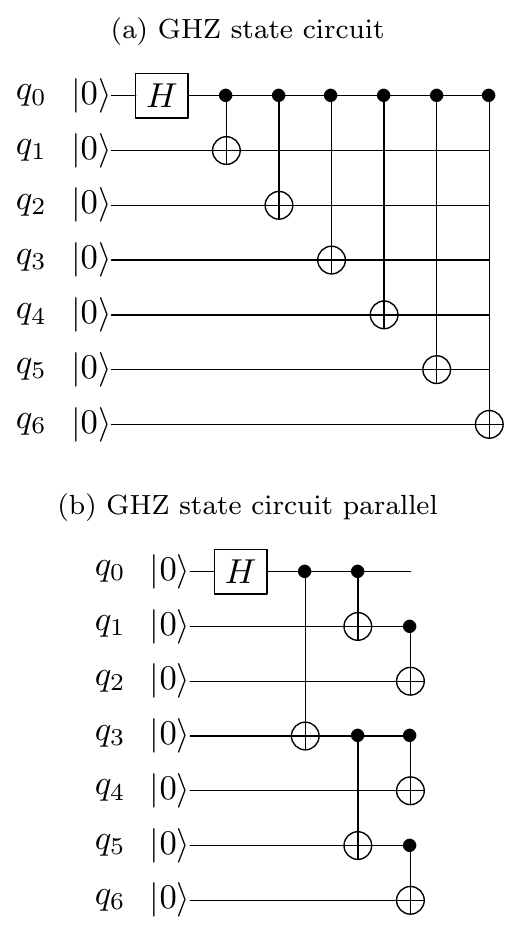}
	\caption{Example preparation circuits for 7-qubit GHZ states. Whenever a CNOT is applied within these circuits, the GHZ state grows in size by 1 qubit. \textbf{(a)} An inefficient embedding where all CNOTs are applied from the primary qubit, resulting in a CNOT depth of 6. \textbf{(b)} An optimal embedding where CNOTs are applied in parallel from qubits that are already included in the growing GHZ state, resulting in a CNOT depth of 3. When constructing GHZ states, we aim to prepare them using a similar embedding to this optimal one modulo the qubit layout of the quantum device.} \label{fig:ghz-circuits}
\end{figure}
GHZ states are highly entangled states. They can be prepared in gate-based quantum devices by initialising a single primary qubit to the $|+\rangle$ state and the other qubits to the $|0\rangle$ state, then CNOT gates are iteratively applied from the primary qubit (or any other qubit that has already had a CNOT applied in this manner) to each other qubit involved in the state, as shown in Figure~\ref{fig:ghz-circuits}. A GHZ state can be expressed as
\begin{equation}
    |\text{GHZ}_{N}\rangle = \frac{|0\rangle^{\otimes N} + |1\rangle^{\otimes N}}{\sqrt{2}},
\end{equation}
where $N$ is the number of qubits in the state. This state has a convenient structure for detecting genuine multipartite entanglement since its density matrix only consists of non-zero entries in each of the the four corners. The GHZ fidelity, $F$, can be calculated by measuring two observables \textit{population} $P$ and \textit{coherence} $C$ as
\begin{equation}\label{eq:fidelity_p_plus_c}
    F := \frac{P + C}{2},
\end{equation}
where $P = \rho_{00\ldots 0, 00\ldots 0} + \rho_{11\ldots 1, 11\ldots 1}$ can be directly measured as the GHZ populations and $C = |\rho_{11\ldots 1, 00\ldots 0}| + |\rho_{00\ldots 0, 11\ldots 1}|$ can be measured through parity oscillations~\cite{leibfried2005creation, monz201114}, or multiple quantum coherences (MQC)~\cite{wei2020verifying, baum1985multiple, wei2018exploring, garttner2017measuring}. A GHZ fidelity greater than 0.5 is sufficient to demonstrate that the state exhibits GME~\cite{leibfried2005creation}. The following description of the process for measuring the fidelity is adapted from the details presented in Appendix~\ref{sec:MQC-explanation}. To measure the coherence, we use the method of MQC due to its promising robustness to noise by using a refocusing $\pi$-pulse similar to a Hahn echo~\cite{hahn1950spin} to refocus low frequency noise and reduce dephasing~\cite{wei2020verifying}. Since $X$ gates stabilise the GHZ state, this does not affect fidelity computations, and the usual second pulse used to cancel the first in refocusing pulse sequences can be omitted. The coherence can be calculated by measuring the overlap signals $S_\phi = \text{Tr}(\rho_\phi \rho)$, where $\rho_\phi = e^{-i\frac{\phi}{2}\sum_j \sigma_z^j} \rho  e^{i\frac{\phi}{2}\sum_j \sigma_z^j}$ is produced by rotating each qubit of the state~$\rho$ by~$\phi$ about the~$Z$-axis. Essentially, $S_\phi$ is the probability of measuring the zero state after encoding the GHZ state, applying the $\phi$ phase rotations, then decoding the GHZ state. For an ideal GHZ state, the phase produced by rotating each individual qubit accumulates and is equivalent to adding a phase of $N\phi$ to the state, i.e. $\frac{1}{\sqrt{2}}(|00\ldots 0\rangle + e^{-iN\phi} |11\ldots 1\rangle)$, which reduces the overlap signal to
\begin{equation}
    S_\phi^{\text{ideal}} = \frac{1}{2}[1 + \cos (N\phi)],
\end{equation}
where the constant term corresponds to the diagonal elements of the density matrix and the cosine term corresponds to the off-diagonal corner elements. By Fourier transforming $S_\phi$ we can obtain the MQC amplitudes
\begin{equation}
    I_q = \mathcal{N}^{-1}|\sum_\phi e^{i q \phi} S_\phi|,
\end{equation}
where $\phi = \frac{\pi j}{N+1}$ for $j = 0, 1, \ldots, 2N+1$ to detect up to frequency $N+1$ and $\mathcal{N} = 2N + 2$ is the number of angles $\phi$ in the summation.
The coherence is then calculated as $C = 2\sqrt{I_N}$.
There is a slow free rotation that occurs in idle qubits throughout the computation. To help alleviate these drift effects, a refocusing $\pi$-pulse applied between the GHZ encoding and decoding steps is used. The procedure to compute each of the overlap signals $S_\phi$ consists of the following steps.
\begin{itemize}
    \item[1.] Encode the GHZ state by applying the preparation circuit over the desired qubits.
    \item[2.] Apply a refocusing $\pi$-pulse as an $X$ gate on each qubit.
    \item[3.] Apply the phase gate rotation of $\phi$ about the $Z$-basis.
    \item[4.] Decode the GHZ state by applying the inverse of its preparation circuit.
    \item[5.] The overlap signal $S_\phi$ is then the probability of measuring the zero state $|00\ldots 0\rangle$.
\end{itemize}
Further details involved in expressing the fidelity concretely are presented in Appendix~\ref{sec:MQC-explanation}.
\subsection*{Quantum readout-error mitigation (QREM)}
On noisy intermediate-scale quantum (NISQ) hardware, measurement represents one of the largest single-component error sources. This is especially significant in low-depth circuits where the number of qubit readouts is comparable to the total number of gates applied.
Typical readout error rates of a few percent per qubit can quickly scramble the results of any quantum algorithm. 
In particular, they may obfuscate the results of an otherwise relatively well-performing device. 
Fortunately, however, readout error is far more addressable than other quantum errors on QIPs. First, excluding the case of mid-circuit measurement, readout errors do not propagate. Secondly, and more importantly, readout error can be treated as classical error to a good approximation for typical superconducting devices. 
We employ the quantum readout error mitigation (QREM) procedure developed in \cite{maciejewski2020mitigation} in our experiments. \par 
Readout error can be difficult to characterise. Formally, laboratory measurements output estimates of the object $\text{Tr}[E(U\rho U^\dagger)]$, for some noisy state $\rho$, intermediate unitary circuit $U$, and positive operator-valued measure (POVM) elements $E$. Typically, this POVM is the set $\{|0\rangle\langle{0}|,|1\rangle\langle 1|\}$, i.e. measurement in the $Z$-basis.
Any deviation from the ideal case in this object cannot be narrowed down to any one of the three components without further investigation.
Even without intermediate operations, it is impossible to tell where an error falls on the continuum between a faulty state preparation with a perfect measurement, and perfect state preparation with a faulty measurement. \par 
This noise falls under a category termed state preparation and measurement (SPAM) error. 
There are techniques equipped to self-consistently characterise SPAM \cite{sandia-GST}, but are typically computationally costly.
Instead, it often suffices as a good approximation for modern hardware to assume the former end of the spectrum: that is, a device capable of perfectly delivering a $\ket{0}$ state but with some error in measuring it. The magnitude of the different error sources are often two or more orders of magnitude apart, justifying the assumption.
The benefit of this approximation is that the conditional probability distribution $p(\text{measure}\:x|\text{prepared}\:y)$ can be characterised simply by preparing all states $y$, and counting all measurements $x$. 
Whilst the number of preparation states to consider grows exponentially in the number of qubits, it can simplify matters greatly to assume that each faulty POVM element has a tensor product structure. That is, that the readout errors are either local-only or have correlations with limited spatial extent.
In the case of solely local errors, the characterisation can be performed in a constant number of circuits -- by preparing and measuring both $\ket{0}^{\otimes n}$ and $\ket{1}^{\otimes n}$. For limited local connections, this is constant with the number of qubits and exponential in the extent of the correlations. \par
For this work, we operate under the first-order regime of local errors. Each state $\ket{\psi}$ is transformed according to~$\ket{\psi'} = A_\text{local} \ket{\psi}$, where $A_\text{local} := \bigotimes_{i=1}^n A_i$. Each stochastic matrix $A_i$, is called the \emph{calibration matrix} for qubit $i$ and is defined as
\begin{equation}
\label{cal-mat}
    A_i := \begin{pmatrix} p_i(0|0) & p_i(0|1) \\ p_i(1|0) & p_i(1|1) \end{pmatrix}, 
\end{equation}
where the notation $p_i(x|y)$ indicates the probability of measuring the state $\ket{x}$ given the prepared state $\ket{y}$ for qubit $i$. The calibration matrices $A_i$ are constructed using the QDT~\cite{lundeen2009tomography} technique. 
In principle, the QREM procedure consists of collecting a vector of measurement outcomes and multiplying this by $A_\text{local}^{-1}$ in order to obtain the pre-readout state. In practice, however, imperfections in the calibration and the mitigation -- caused at the very least by finite sampling error -- can result in a final probability vector with negative elements. To combat this, a projection method is usually used to find the closest physical probability vector (on the unit simplex, with positive elements summing to 1)~\cite{smolin2012efficient}. To summarise: a stochastic calibration matrix is constructed for each individual qubit; the tensor product of the inverse of each calibration matrix is multiplied by the measured results vector; finally, the closest probability vector is found, representing the best estimate of the quantum state before readout error. In Appendix~\ref{sec:qrem-assumptions}, we estimate the size of the approximations made in performing QREM using this simplified procedure. The maximum difference in findings is small when compared to the error bars of our results. 

The initial measured probability vector can only contain probability values that are at least as large as the equivalent of a single shot, $1/$(shot count), due to the finite number of shots, hence the maximum number of states with non-zero probability is equal to the shot count. However, applying the $A_{\text{local}}^{-1}$ to the probability vector often results in a high number of states having very small non-zero probabilities (sometimes much less than $1/100$ of a shot). The amount of memory required to precisely store a single probability vector after applying $A_{\text{local}}^{-1}$ no longer scales with the number of shots. Instead, the memory scales as $2^N$ for $N$ qubits, since sparse vectors are no longer useful. To help alleviate this computational resource requirement, we apply the inverse calibration matrices for each qubit $A^{-1}_i$ sequentially on the relevant elements of the probability vector. After every application of $A^{-1}_i$, probability values with magnitude below some small threshold are set to zero. A threshold value of $1/10$ of a shot was used, which equates to a probability of just over $1.2 \times 10^{-5}$ for 8192 shot experiments. This value was chosen because it was high enough to considerably reduce the overall computation time, while the approximation remains close to the full calculation. When testing this approximation on a 19-qubit state, the average computation time for each application of QREM dropped from 66 sec to 3.8 sec (performed on a laptop with 2.70 GHz Intel i7-7500U CPU and 16 GB RAM), and the fidelity was found to match the full calculation by up to five decimal places. The reduced computation time enabled each sample from the 27-qubit GHZ state to be computed within two days.

\subsection*{Parity Verification}
We examine the extent to which the fidelity of GHZ states can be improved through error-detecting state preparation techniques. 
In particular, we employ the parity-checking technique used in \cite{Gottesman2016} for the $[[4,2,2]]$ code. 
The goal here is to extract information about errors occurring within the GHZ state without disturbing the state itself. 
This can be achieved with a parity check between qubits of the state, where two CNOTs (control ($c$) $\rightarrow$ target ($t$), denoted CNOT$^c_t$) are performed which are controlled on GHZ state qubits and target a single ancilla qubit initialised in the ground state.
In the case of an even number of bit-flip errors, parity-checking will leave the state of the ancilla qubit unchanged. Alternatively, for an odd number of bit-flips, the state of the ancilla will flip, signalling a detectable error within the GHZ state. More explicitly, a parity check on qubits $q_1$ and $q_2$ using ancilla qubit $a$ will result in the transformations
\begin{align*}
    \ket{00}_{q_1,q_2}\ket{0}_{a} \xrightarrow[]{\text{CNOT$^{q_1}_{a}$CNOT$^{q_2}_{a}$}}\ket{00}_{q_1,q_2}\ket{0}_{a},\\
    \ket{01}_{q_1,q_2}\ket{0}_{a} \xrightarrow[]{\text{CNOT$^{q_1}_{a}$CNOT$^{q_2}_{a}$}}\ket{01}_{q_1,q_2}\ket{1}_{a},\\
    \ket{10}_{q_1,q_2}\ket{0}_{a} \xrightarrow[]{\text{CNOT$^{q_1}_{a}$CNOT$^{q_2}_{a}$}}\ket{10}_{q_1,q_2}\ket{1}_{a},\\
    \ket{11}_{q_1,q_2}\ket{0}_{a} \xrightarrow[]{\text{CNOT$^{q_1}_{a}$CNOT$^{q_2}_{a}$}}\ket{11}_{q_1,q_2}\ket{0}_{a}.
\end{align*}
Measuring the ancilla in the $\ket{1}_a$ state implies a bit-flip error has occurred within the GHZ state, allowing the erroneous shot measurement to be discarded from the results. In the case of measuring the $\ket{11}_{q_1, q_2}\ket{0}_a$ state, the two bit-flip errors are not detected. However, if $p$ is the single-qubit probability of a bit-flip occurring for qubits $q_1$ and $q_2$, then the probability of bit-flips occurring on both $q_1$ and $q_2$ is small $p^2$. For example, assuming bit-flip probabilities are moderate at $p=0.25$, if $\ket{0}_a$ is already measured (assuming no readout errors), then there is a probability of $(1-p)^2/((1-p)^2+p^2) = 0.9$ that the state has no bit-flip errors $\ket{00}_{q_1, q_2}$ and a probability of $p^2/((1-p)^2+p^2)=0.1$ that the state has two bit-flip errors $\ket{11}_{q_1, q_2}$.

Under the typical framework of quantum error correction, the syndrome measurement of an ancilla qubit will be fed forward to correct the original state. Since error correction is not achievable within the current restrictions of quantum hardware, we instead focus on detection so that runs can be post-selected. This is usually seen as the condition under which fault-tolerance can be demonstrated on near-term quantum devices~\cite{Gottesman2016}.
\subsection*{Fidelity of GHZ states with parity verification in the \textit{ibmq\_montreal} device}
\begin{figure*}[tbh]
	\centering
     \includegraphics[scale=1]{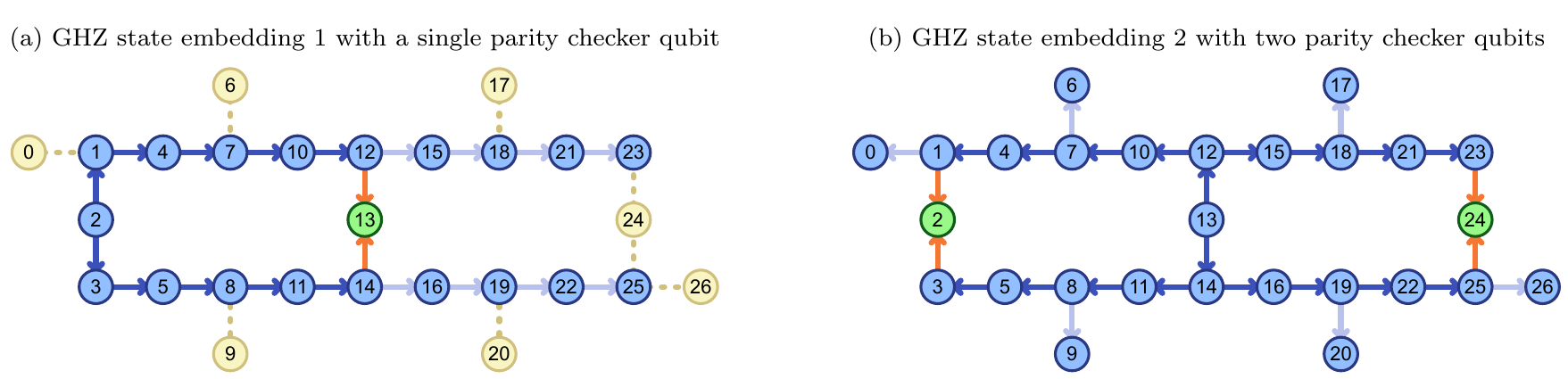}
	\caption{Diagrams showing how the GHZ states are constructed on the \textit{ibmq\_montreal} device. Blue vertices represent qubits of the GHZ state, green vertices represent parity checking qubits, and lighter beige vertices represent qubits that are not directly involved in the experiment. Arrows represent the direction of CNOTs (control ($c$) $\rightarrow$ target ($t$), denoted CNOT$^c_t$). Dark blue arrows indicate CNOTs used for constructing the GHZ state for all state sizes, light blue arrows indicate CNOTs growing the GHZ state, orange arrows indicate CNOTs used for parity verification, dotted beige lines indicate that no two-qubit gates are applied. \textbf{(a)} Shows the construction of the state sizes 11 to 19 beginning at qubit 2 and including the single parity checking qubit 13. The CNOT$^2_1$ gate is performed before the CNOT$^2_3$ gate. The states are incrementally grown from the initial 11 qubits by including qubits in the order: 15, 16, 18, 19, 21, 22, 23, and 25. \textbf{(b)} Shows the construction of the state sizes 19 to 25 beginning at qubit 13 and including the two parity checking qubits 2 and 24. The initial CNOTs are performed in the order CNOT$^{13}_{12}$, CNOT$^{13}_{14}$, CNOT$^{12}_{10}$, CNOT$^{12}_{15}$, CNOT$^{14}_{11}$, and CNOT$^{14}_{16}$. The states are incrementally grown from the initial 19 qubits by including qubits in the order: 9, 6, 20, 17, 0 and 26. The 26 and~27-qubit states do not use qubits 2 and 24 as parity checking qubits, instead the 26-qubit GHZ state includes qubit 2 with a CNOT$^{3}_{2}$ in its construction while the~27-qubit state additionally includes qubit 24 with a CNOT$^{23}_{24}$ gate.} \label{fig:ghz-embeddings}
\end{figure*}
\begin{figure*}
     \centering
     \includegraphics[scale=1]{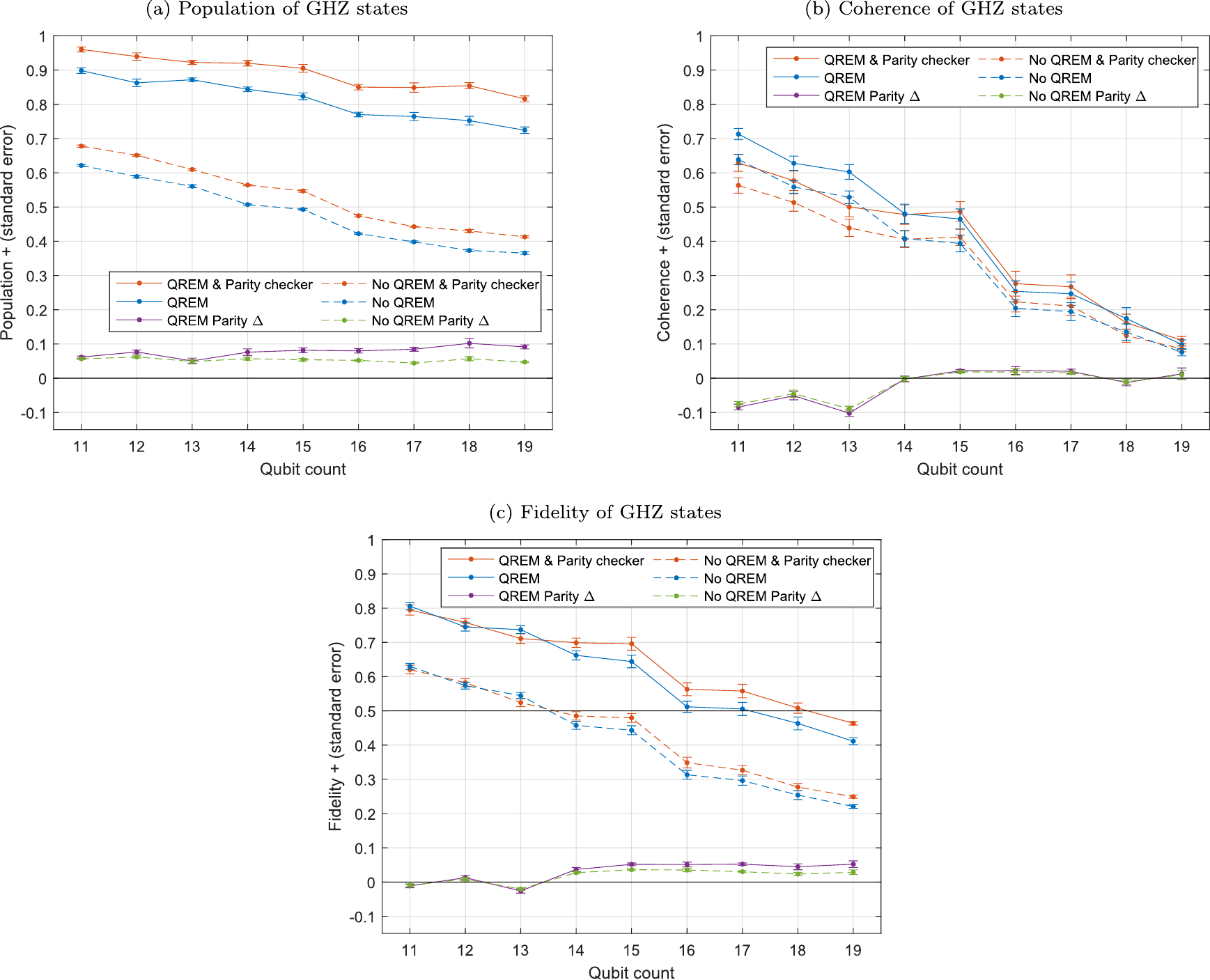}
        \caption{\textbf{Embedding 1}. Measured observables for GHZ states varying in sizes from 11 to 19 qubits on the \emph{ibmq\_montreal} device. These states are prepared using the embedding shown in Fig.~\ref{fig:ghz-embeddings}a. Each plot compares combinations of applying QREM and using parity-checker qubit 13 to verify the equality of states of qubits 12 and 14. The parity $\Delta$ refers to the difference between using and not using parity verification. The parity-checking qubit requires the CNOT circuit depth to increase by one for state sizes 11 to 14 while the depth is unchanged for state sizes 15 to 19 since the parity-checker CNOTs are performed in parallel. The standard errors were calculated from 8 independent runs of the experiment where all circuits were performed using 8192 shots. The corresponding CNOT circuit depths and counts are displayed in Table~\ref{tab:CNOT-depth-and-counts}.  \textbf{(a)} The population is calculated as the summed occupancies of measured states $|00\ldots 0\rangle$ and $|11 \ldots 1\rangle$ from the prepared state. Parity verification is shown to significantly increase the population values for all state sizes. \textbf{(b)} The coherence is calculated using the~QMC method. The coherence appears to be significantly decreased by parity verification for state sizes 11 to 13. This is likely due to the increased CNOT circuit depth required for parity verification on states of sizes 11 to 14. The coherence appears to be mostly unchanged for state sizes from 14 to 19 qubits. \textbf{(c)} The fidelity is calculated as the average over the population and coherence. When using the parity check, the average fidelity over qubit sizes 15 to 19 (where the CNOT depth is unchanged) increases by $0.051 \pm 0.007$ with QREM and $0.031\pm 0.004$ without QREM. }\label{fig:fidelity-population-coherence-ghz}
\end{figure*}

\begin{table*}
\begin{tabular}{|c|cc|cc|cc|cc|}
\cline{2-9}
\multicolumn{1}{ c| }{} & \multicolumn{2}{ c| }{Population} & \multicolumn{2}{ c| }{Population (parity)} & \multicolumn{2}{ c| }{Coherence} & \multicolumn{2}{ c| }{Coherence (parity)} \\
\cline{1-9}
\multicolumn{1}{ |c| }{State size (qubits)} & Depth & Count & Depth & Count & Depth & Count & Depth & Count \\
\cline{1-9}
\multicolumn{9}{ |c| }{Embedding 1 -- GHZ state sizes 11 to 19 (one parity-checker qubit)}\\
\hline
11 & 6 & 10 & 7 & 12 & 12 & 20 & 13 & 22\\
12 & 6 & 11 & 7 & 13 & 12 & 22 & 13 & 24\\
13 & 7 & 12 & 8 & 14 & 14 & 24 & 15 & 26\\
14 & 7 & 13 & 8 & 15 & 14 & 26 & 15 & 28\\
15 & 8 & 14 & 8 & 16 & 16 & 28 & 16 & 30\\
16 & 8 & 15 & 8 & 17 & 16 & 30 & 16 & 32\\
17 & 9 & 16 & 9 & 18 & 18 & 32 & 18 & 34\\
18 & 9 & 17 & 9 & 19 & 18 & 34 & 18 & 36\\
19 & 10 & 18 & 10 & 20 & 20 & 36 & 20 & 38\\
\cline{1-9}
\multicolumn{9}{ |c| }{Embedding 2 -- GHZ state sizes 19 to 27 (two parity-checker qubits)}\\
\cline{1-9}
19 & 7 & 18 & 8 & 22 & 14 & 36 & 15 & 40\\
20 & 7 & 19 & 8 & 23 & 14 & 38 & 15 & 42\\
21 & 7 & 20 & 8 & 24 & 14 & 40 & 15 & 44\\
22 & 7 & 21 & 8 & 25 & 14 & 42 & 15 & 46\\
23 & 7 & 22 & 8 & 26 & 14 & 44 & 15 & 48\\
24 & 7 & 23 & 9 & 27 & 14 & 46 & 16 & 50\\
25 & 7 & 24 & 9 & 28 & 14 & 48 & 16 & 52\\
26 & 7 & 25 & - & - & 14 & 50 & - & -\\
27 & 7 & 26 & - & - & 14 & 52 & - & -\\
\hline
\end{tabular}
\caption{CNOT circuit depths and counts required to perform the corresponding experiments.\label{tab:CNOT-depth-and-counts} }
\end{table*}

\begin{figure*}
     \centering
    \includegraphics[scale=1]{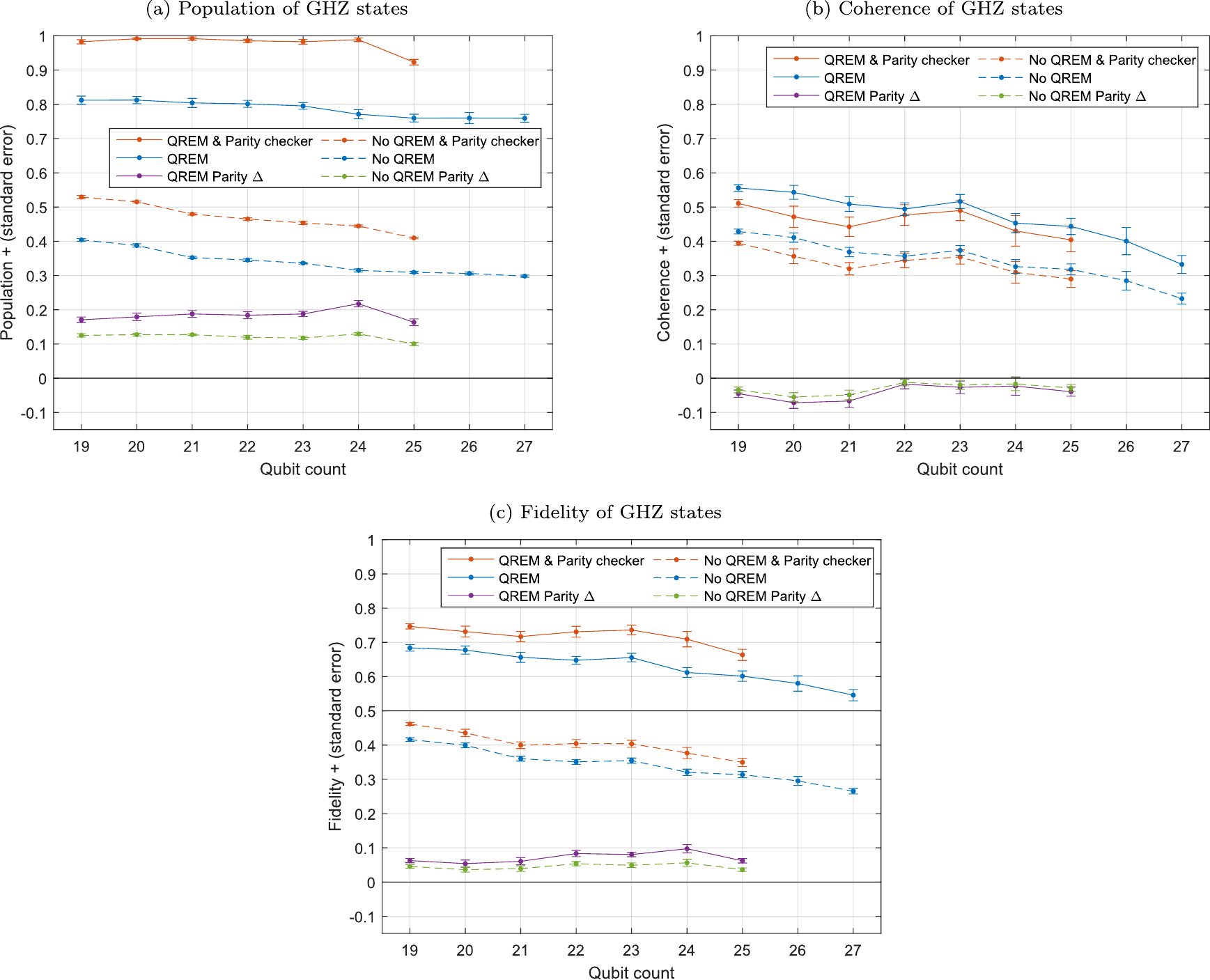}
    \caption{\textbf{Embedding 2}. Measured observables for GHZ states varying in sizes from 19 to 27 qubits on the \textit{ibmq\_montreal} device. These states are prepared using the more efficient embedding shown in Fig.~\ref{fig:ghz-embeddings}b lowering CNOT circuit depths resulting in higher fidelity values for large states than what is shown in Figure~\ref{fig:fidelity-population-coherence-ghz}. Each plot compares combinations of applying quantum readout-error mitigation (QREM) and using parity-checker qubits 2 and 24 to verify the equality of states of qubits 1 and 3, and qubits 23 and 25 respectively. The parity $\Delta$ refers to the the difference between using and not using parity verification. The preparation of each state without parity verification requires the same CNOT circuit depth since the additional qubits can be included by applying CNOT gates in parallel. When including the two parity-checking qubits, the CNOT circuit depth increases by one for sizes 19 to 23 and increase by two for sizes 24 and 25. The standard errors were calculated from 8 independent runs of the experiment where all circuits were performed using 8192 shots. All corresponding CNOT circuit depths and counts are displayed in Table~\ref{tab:CNOT-depth-and-counts}. \textbf{(a)} The population is calculated as the probability of obtaining $|00\ldots 0\rangle$ or $|11\ldots 1\rangle$ upon measurement of the prepared GHZ state. Parity verification using the two parity-checking qubits is shown to significantly increase the population values for all state sizes from 19 to 25 qubits. \textbf{(b)} The coherence is calculated using the quantum multiple coherences (QMC) method. No increase in coherence is observed using parity verification for state sizes 19 to 25. For state sizes 19 to 21 in particular, the measured coherence decreases slightly. \textbf{(c)} The fidelity is calculated as the average over the population and coherence. When using parity verification, the fidelity increases on average by $0.072 \pm 0.009$ with QREM and $0.045\pm 0.007$ without QREM over state sizes 19 to 25.}\label{fig:fidelity-population-coherence-ghz-large}
\end{figure*}

\begin{figure*}[!htb]
     \centering
     \includegraphics[scale=1]{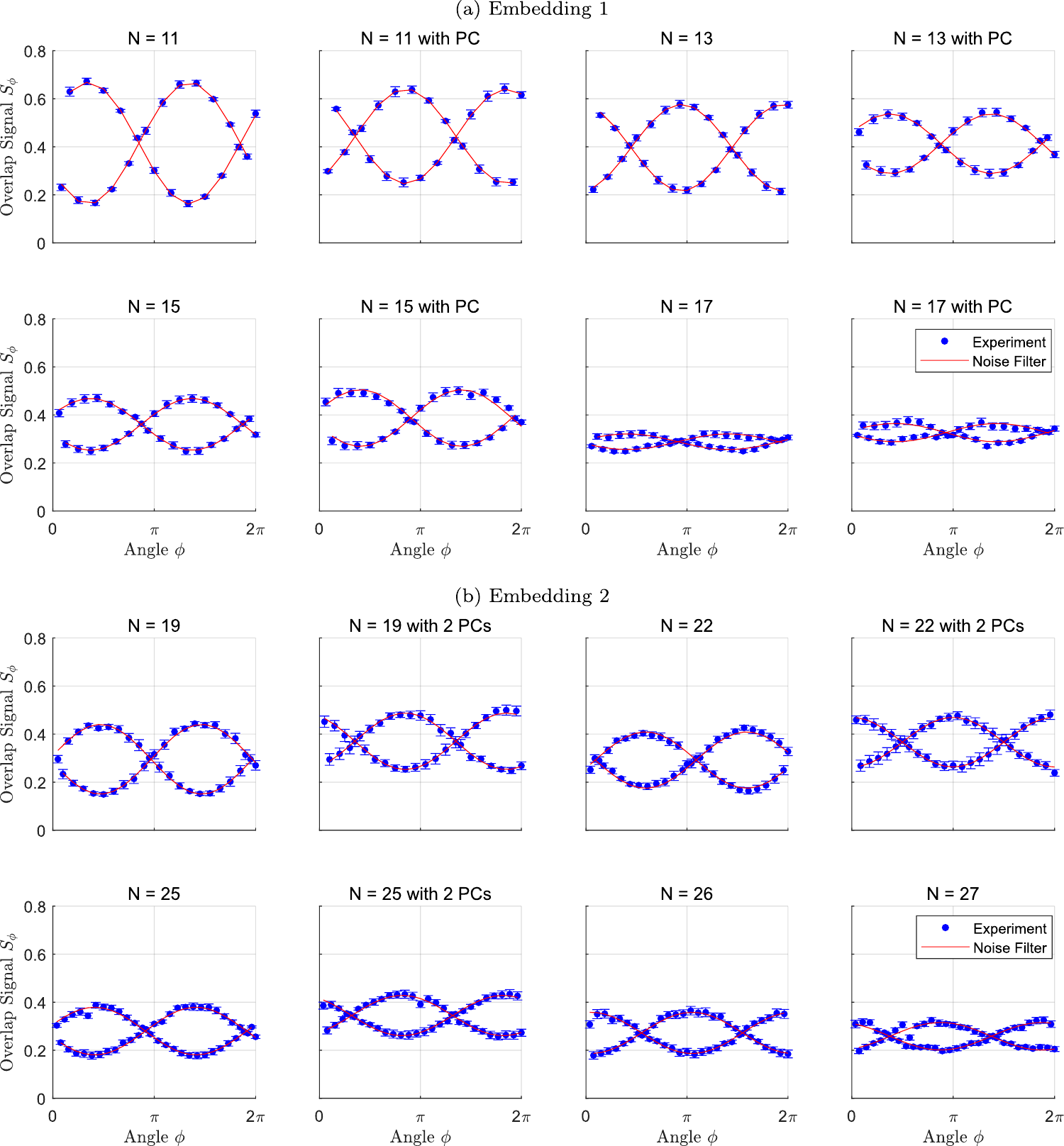}
        \caption{Measured MQC overlap signals for various size GHZ states prepared on the \textit{ibmq\_montreal} quantum device and corrected using QREM. The experimental results and standard errors were calculated from 8 independent runs of the experiment performed on the same device calibration where all circuits were executed using 8192 shots. The noise filtered data are calculated by taking the Fourier transform of the experimental results, zeroing frequencies that are not $0, \pm N$, then inverting the Fourier transform. There is a phase shift present in the measured $S_\phi$ values that is likely caused by free rotation in idle qubits. The refocusing $\pi$-pulse is applied to the encoded GHZ state before decoding to help nullify these qubit drift effects, however the GHZ encoding and decoding operations take slightly different amounts of time to compute due to pulse alignment restrictions in the software. This phase shift is more pronounced when parity verification is applied since the parity-checking CNOTs are performed before the refocusing $\pi$-pulse, offsetting the number of gates before and after refocusing. \textbf{(a)} GHZ state is implemented using embedding 1 shown in Figure~\ref{fig:ghz-embeddings}a. Parity verification uses qubit 13 as a single parity-checker (PC) qubit to verify the states of qubits 12 and 14. \textbf{(b)} GHZ state is implemented using embedding 2 shown in Figure~\ref{fig:ghz-embeddings}b. Parity verification for states of sizes 19 to 25 qubits uses qubits 2 and 24 as parity-checkers to verify the equality of states of qubits 1 and 3, and qubits 23 and 25 respectively.}\label{fig:MQC-22q}
\end{figure*}

In this section, we measure the fidelity on a range of GHZ state sizes prepared on the \emph{ibmq\_montreal} device and observe the effects of using parity-checking qubits to verify them. The following experiments prepare GHZ states and parity verification on \emph{ibmq\_montreal}'s qubit layout via the diagrams shown in Figure~\ref{fig:ghz-embeddings}. The parity verification CNOTs are performed directly after preparing the GHZ states. In the case of measuring the coherence, these CNOTs are in the middle of the circuit, since there is a GHZ decoding sequence of gates applied afterwards. Ideally the parity qubits would be measured immediately after their corresponding CNOTs have been applied to help reduce parity qubit error and abandon the computation as soon as possible in the case of an error, however superconducting devices do not yet robustly support this functionality and hence the parity qubits are measured at the end of the computation along with the other qubits. The population is calculated by summing the measured $00 \ldots 0$ and $11 \ldots 1$ occupancies from the prepared GHZ states and the coherence is calculated by using the MQC method. Additionally, QREM was used on GHZ state qubits to reduce the effects of noise due to measurement errors. The parity-checking qubit cannot have QREM applied because it is a single shot measurement that directly affects whether the shot is discarded. For all of the experiments, the standard errors were calculated from 8 independent runs of the experiment where all circuits were performed using 8192 shots. All computations were performed on the Spartan high performance computing system~\cite{meade_lafayette_sauter_tosello_2017}. The corresponding CNOT circuit depths and counts for each experiment are shown in Table~\ref{tab:CNOT-depth-and-counts}.

GHZ states of incrementally increasing sizes from~11 to~19 qubits were prepared using the embedding shown in Figure~\ref{fig:ghz-embeddings}a. The smallest state of 11 qubits is the minimum size required to include a parity-checker qubit within the \textit{ibmq\_montreal} device's layout. The initial Hadamard is performed on qubit 2 and CNOTs grow the state in the up-right and down-right directions. The parity qubit is at location 13 and was added to the circuits by applying two CNOTs with control qubits 12 and 14 and target qubit 13 directly after preparing the GHZ states. The resulting fidelity, population and coherence measurements were obtained within the same device calibration and are shown in Fig.~\ref{fig:fidelity-population-coherence-ghz}, the overlap signals for state sizes 11, 13, 15, and 17 for the calculation of the coherences are shown in Figure~\ref{fig:MQC-22q}a. Performing the parity-checking verification requires the CNOT circuit depth to increase by one for states of size 11 to 14, while the depth is unchanged for states of size 15 to 19 since the parity-checker CNOTs are performed in parallel to the preparation circuit. When using the parity check, there is a noticeable improvement to fidelity for state sizes that do not require the CNOT circuit depth to increase, namely for state sizes 15 to 19, where the average fidelity increases by $0.051\pm 0.007$ with QREM and $0.031\pm 0.004$ without QREM. The population values gain a consistent advantage through parity verification, while the coherence values appear mostly unchanged for sizes 15 to 19 and slightly decrease for sizes 11 to 14. The decrease in coherence is likely due to the inclusion of the additional CNOTs increasing the CNOT circuit depth and introducing noise. Measurement error on the parity-checking qubit could also be a contributing factor by resulting in a small number of the wrong circuits being discarded. We also observe that for these prepared states, the measured population and its change due to QREM are much larger than for the coherence. These observations are expected because dephasing is the dominant error channel in superconducting qubits, rather than population errors. Therefore we expect the coherence to be the limiting factor in generating these states to high fidelity. The effects of QREM on population are likely more drastic in their improvement than with the coherence because readout errors are significantly more likely to occur from $\ket{1}$ to $\ket{0}$ compared to the other way around. This leads to population measurements being far more sensitive to measurement error than the coherence curve (since the occupancies of both $\ket{00\ldots 0}$ and $\ket{11\ldots 1}$ are required to measure the population while only the occupancy of $\ket{00\ldots 0}$ is required for the coherence).

To further explore the effects of parity verification on larger states, GHZ states were prepared for incrementally increasing sizes from 19 to 25, with and without two parity-checking qubits. The \textit{ibmq\_montreal} device has a convenient layout for preparing larger GHZ states, since it allows a relatively high number of branches when constructing them, allowing more CNOTs to be computed in parallel as shown in Figure~\ref{fig:ghz-embeddings}b. The initial Hadamard is performed on qubit 13 and CNOTs grow the state in the four directions up-left, up-right, down-right, down-left. The parity-checking qubits are located at 2 and 24 and were added to the circuits by applying, directly after preparing the GHZ states, four CNOTs, two with control qubits 1 and 3 and target qubit 2, and two with control qubits 23 and 25 and target qubit 24. Due to the parallel nature of this state embedding, each of the preparation circuits for sizes 19 to 25 have equal CNOT circuit depths, with depth increasing by one when using parity verification for sizes 19 to 23 and increasing by two for sizes 24 and 25. The resulting fidelity, population and coherence measurements were obtained within the same device calibration and are shown in Figure~\ref{fig:fidelity-population-coherence-ghz-large}, the overlap signals for state sizes 19, 22, and 25 are shown in Figure~\ref{fig:MQC-22q}b. When using parity verification, the fidelity over state sizes 19 to 25 increases on average by $0.072\pm 0.009$ with QREM and $0.045\pm 0.007$ without QREM. In particular, for the 25-qubit GHZ states, fidelities of $0.664 \pm 0.016$ and $0.601\pm 0.015$ were measured using QREM with and without parity verification respectively. The measured population values are consistently higher for all state sizes, while the measured coherence values show no increase in any of the state sizes and in particular, the coherence values decrease slightly for state sizes 19 to 21 qubits.

To test the extent of possible GME within the device, GHZ states of sizes 26 and 27 qubits were additionally prepared on the device (with embedding shown in Figure~\ref{fig:ghz-embeddings}b). The experiments were performed on the same calibration as the previous experiments relating to the GHZ state sizes of 19 to 25 and the results are shown in Figure~\ref{fig:fidelity-population-coherence-ghz-large}, with corresponding overlap signals for the calculation of coherence shown in Figure~\ref{fig:MQC-22q}b. The measured fidelities using QREM were found to be $0.580\pm 0.022$ and $0.546\pm 0.017$ for the 26 and 27-qubit states respectively. These fidelities are above the 0.5 threshold with confidence levels of 99.6\% and 98.6\% respectively, thus GME was detected in both states.

We suspect it is possible to extend parity verification to consist of three or more qubits with majority rules, instead of a single qubit, to reduce measurement error. This could be beneficial because current measurement error mitigation techniques cannot be applied to parity qubits due to the parity being a single shot measurement. However, it may not be worth it due to introduced CNOT errors potentially being larger than the original measurement error.

\section*{Discussion}
A variety of GHZ states were prepared on the 27-qubit \textit{ibmq\_montreal} device and the GHZ fidelities, populations and coherences were measured. We report the preparation of a 27-qubit state with a GHZ fidelity of $0.546\pm 0.017$, demonstrating GME across the full device with a confidence level of 98.6\%. This is currently the largest state prepared on a quantum system that has been reported to exhibit GME. We further benchmarked the effects of parity verification on the fidelity, population and coherence. Two GHZ state preparation embeddings on the physical qubit layout were used. The first embedding generated GHZ states of incrementally increasing sizes from 11 to 19 qubits while using a single parity-checking qubit. When the parity-checker qubits did not require the CNOT depth of the circuit to increase, which was the case for state sizes 15 to 19, the fidelity increased by an average of $0.051\pm 0.007$ with QREM and $0.031\pm 0.004$ without QREM. The fidelity was mostly unchanged for sizes 11 to 13 and was higher for size 14. The lack of fidelity improvement for state sizes 11 to 13 was most likely due to the noise introduced by the CNOTs of the parity-checking qubit and their implementation resulting in the CNOT circuit depth increasing by one for state sizes 11 to 14. The population values increased for all state sizes when the parity-checker qubit was used, while the coherence values appear mostly unchanged for state sizes from 15 to 19 and slightly decrease for sizes from 11 to 14. The second embedding generated GHZ states of incrementally increasing sizes of 19 to 25 qubits while using two parity-checker qubits. It utilised the layout of the \textit{ibmq\_montreal} device by performing as many CNOTs as possible in parallel to minimise the CNOT circuit depth. The parity-checking qubits increased the CNOT circuit depth for state sizes 19 to 23 by one and for sizes 24 and 25 by two. We observed that the parity-checking procedure with two qubits increased the fidelity slightly for all state sizes resulting in an average increase of $0.072\pm 0.009$ with QREM and $0.045\pm 0.007$ without QREM. In particular, the 25-qubit state, which was the largest state tested using parity checkers, had its fidelity increase from $0.601\pm 0.015$ to $0.664\pm 0.016$ with QREM. The population values consistently increased for all state sizes, while the coherence values decreased slightly for state sizes 19, 20, 21 and 25 qubits and were unchanged for state sizes 22, 23 and 24.
The results show that the effect of parity verification led to a detectable improvement of GHZ fidelity, although relatively modest on current IBM Quantum hardware.

\section*{Acknowledgements}
This work was supported by the University of Melbourne through the establishment of an IBM Q Network Hub at the University. CDH is supported by a research grant from the Laby Foundation.

\section*{Author contributions statement}
GJM, GALW, CDH and LCLH conceived the experiments. GJM obtained and processed the main results, while GALW obtained and processed the QREM analysis of Appendix A. GJM and GALW wrote the manuscript with input from CDH and LCLH.

\section*{Data availability}

The datasets generated and/or analysed during the current study are available from the corresponding author on reasonable request.

\section*{Additional information}

\emph{Non-financial competing interests}: The authors are supported by the University of Melbourne through the establishment of an IBM Q Network Hub at the University.

\bibliographystyle{unsrt}

\section*{Appendix}
\appendix
\renewcommand\thefigure{\thesection.\arabic{figure}}
\setcounter{figure}{0} 
\renewcommand\thedefinition{\thesection.\arabic{definition}}
\setcounter{definition}{0} 
\renewcommand\thelemma{\thesection.\arabic{lemma}}
\setcounter{lemma}{0}
\renewcommand\thetheorem{\thesection.\arabic{theorem}}
\setcounter{theorem}{0}
\renewcommand\thecorollary{\thesection.\arabic{corollary}}
\setcounter{corollary}{0}

\section{Examining QREM assumptions on device hardware}\label{sec:qrem-assumptions}
In the main text, we briefly outlined some simplifying assumptions made in order to implement quantum error mitigation more efficiently. Respectively, these were: measurement error is significantly larger than preparation error, measurement error is predominantly classical, and measurement error is uncorrelated between qubits. Here, we examine these assumptions on chosen subsets of the device. To this end, we employ gate set tomography (GST) for characterisation. GST is a self-consistent extension of regular quantum process tomography (QPT) which includes the SPAM operations in its maximum likelihood estimation. A characterisation of the $i$th qubit is performed across an entire gate set, $\{\rho_i^0, E_i^0, E_i^1, G_i^0, G_i^1,\cdots\}$, whereby the preparation and measurement operators are also estimated. Importantly, this allows us to robustly evaluate the effects of the liberties we have taken. 
For an overview of GST and its implementation, see \cite{sandia-GST,White2019,intro-GST}. We use the well-known quantum characterisation, verification and validation (QCVV) Python package \texttt{PyGSTi} for this part of the implementation \cite{pygsti-arxiv}. \par 
Testing these assumptions takes place in three parts. First, we perform GST across a range of qubits to obtain an estimate for $\rho_i^0$, $E_i^0$, and $E_i^1$, where $\rho_i^0$ is the initial density matrix, and $E_i^m$ is the POVM element corresponding to when the detector clicks with the result $m$. Next, using these results, we compare the magnitudes of errors in $\rho_i^0$ with $E_i^m$. We also determine the off-diagonals of $E_i^m$; non-zero values correspond to coherence in the POVM effect, which is a strictly non-classical error. These values roughly represent the information discarded when the decision is made to use the stochastic calibration matrix (Eq.~\ref{cal-mat}) rather than the complete POVM.
From the POVM, we construct the calibration matrix $A_i^{\text{POVM}}$. This has the same definition as $A_i$ in Equation~\ref{cal-mat}, except its values are obtained with a more comprehensive procedure. We compare these matrices with the end result obtained from the simpler method in the main text. A similar investigation into the rigours of QREM has been conducted in \cite{Geller_2020}. Ideally, the GST estimates for the noisy probes would be used as the basis for QREM, however for our purposes this would have greatly increased the experimental overhead.
The number of experiments required for the chosen GST characterisation of a single qubit is 550, whereas the prepare-and-measure path takes only two circuits with no post-processing. \par 
Finally, having validated the use of the simpler, cheaper stochastic matrix, we examine the assumption of locality of the measurement error. To do this, we select sets of four qubits. We construct the full calibration matrix across the four qubits -- i.e. by preparing $\{\ket{i}\}_{i=0}^{15}$ and measuring. We then compare this to the tensor product of the four local calibration matrices. We note in hindsight that there have been recent developments for characterising correlated measurement error in an efficient manner \cite{Geller_correlated_QREM}. Although our results do not suggest that the tensor product structure significantly impacts the outcome, this could be a more desirable method in future. \par 

\subsection{SPAM magnitudes and the calibration matrix}
Across the device, we performed GST on six different qubits. From this, we obtain estimates of the initial density matrix $\rho_i^0$, and the two-outcome POVM $\{|0\rangle\langle0|,|1\rangle\langle1|\}$. We also obtain estimates for the gates that comprise the process, but these are not relevant to the discussion. Let $\bar{E}_i^m$ denote one of the noisy two-outcome effect operations for the $i$th qubit. Our aim is firstly to determine how classical the error on $\bar{E}_i^m$ is, as well as estimating the size of the error made in assuming perfect state preparation. The latter is not necessarily as simple as finding the size of the preparation error. Rather, we use our estimates $\bar{E}_i^m$ to determine the calibration matrix that we \emph{would} obtain if state preparation were perfect. Using the Born rule, the elements of this matrix $A_i^{\text{POVM}}$ are:
\begin{equation}
    \begin{split}
        A^{\text{POVM}}_{i\:(0,0)} = p_i(0|0) &= \text{Tr}\left[\bar{E}_i^0|0\rangle\langle0|\right] = \bar{E}^0_{i\:(0,0)},\\
        A^{\text{POVM}}_{i\:(0,1)} = p_i(0|1) &= \text{Tr}\left[\bar{E}_i^0|1\rangle\langle1|\right] = \bar{E}^0_{i\:(1,1)},\\
        A^{\text{POVM}}_{i\:(1,0)} = p_i(1|0) &= \text{Tr}\left[\bar{E}_i^1|0\rangle\langle0|\right] = \bar{E}^1_{i\:(0,0)},\\
        A^{\text{POVM}}_{i\:(1,1)} = p_i(1|1) &= \text{Tr}\left[\bar{E}_i^1|1\rangle\langle1|\right] = \bar{E}^1_{i\:(1,1)}.\\
    \end{split}
\end{equation}

A subtlety in this practice is that $\bar{\rho}^0$ and $\bar{E}_i^m$ are not uniquely fixed by any possible experiment; a type of gauge freedom exists in the estimate. Let $\Lambda(\cdot)$ be some error channel acting on either SPAM operation, and let $G$ be some gate sequence. If $\Lambda$ commutes with $G$ then we have
\begin{equation}
    p_i^m = \text{Tr}\left[G\bar{\rho}^0_i G^\dagger \cdot \Lambda(\bar{E}_i^m)\right] = \text{Tr}\left[G\Lambda(\bar{\rho}_0) G^\dagger \cdot \bar{E}_i^m\right].
\end{equation}
That is, there is a class of estimates for $\rho_i^0$ and $E_i^m$ which are completely physically equivalent under commuting transformations. An error map which fits this description is the depolarising channel $\mathcal{E}_p:$
\begin{equation}
    \mathcal{E}_p(\rho) = (1-p)\rho + \frac{p}{2}\mathbb{1}.
\end{equation}
Therefore, in comparing the GST-produced calibration matrix $A_i^{\text{POVM}}$ with $A_i$, we search all variations of
\begin{equation}
\begin{split}
    E_i^0 &\mapsto (1-p_i)E_i^0 + \frac{p_i}{2}\mathbb{1},\\
    E_i^1 &\mapsto (1-p_i)E_i^1 + \frac{p_i}{2}\mathbb{1},\text{ and}\\
    \rho_i^0&\mapsto \rho_i^0 + p_i|0\rangle\langle0| - \frac{p_i}{2}\mathbb{1}
\end{split}
\end{equation}
which preserve the positivity of $\rho_i^0$. Since these representations of the SPAM operations are physically indistinguishable, we can determine the closest $A_i^{\text{POVM}}$ to $A_i$ for each qubit $i$. The remaining difference between the two will estimate the error produced in constructing the calibration matrix using the simpler methods of the main text. For qubits 1, 3, 12, 14, 23, and 25, we summarise the results of this process in Table~\ref{tab:spam-table}. 
\begin{table*}[]
\centering
\resizebox{\textwidth}{!}{%
\begin{tabular}{@{}lccccc@{}}
\toprule
Qubit \# & $\bar{E}_i^0$                                                                                & $A_i^{\text{POVM}}$                                                              & $A_i$                                                                & $||A_i-A_i^{\text{POVM}}||_2$ & $\frac{1}{2}\left(|\bar{E}^0_{i\:(0,1)}| + |\bar{E}^0_{i\:(1,0)}|\right)$ \\ \midrule
1        & $\begin{pmatrix} 0.9913 & -0.0046 + 0.0255 i \\ -0.0046 -0.0255 i & 0.0630 \end{pmatrix}$  & $\begin{pmatrix} 0.9913 & 0.0630 \\ 0.0087 & 0.9370 \end{pmatrix}$ & $\begin{pmatrix} 0.9803 & 0.0509 \\ 0.0197 & 0.9491 \end{pmatrix}$ & 0.0231        & 0.0046                                                              \\

3        & $\begin{pmatrix} 0.9964 & -0.0002  -0.0024 i \\ -0.0002 + 0.0024 i & 0.0093 \end{pmatrix}$ & $\begin{pmatrix} 0.9964 & 0.0093 \\ 0.0036 & 0.9907 \end{pmatrix}$ & $\begin{pmatrix} 0.9945 & 0.0123 \\ 0.0055 & 0.9877 \end{pmatrix}$ & 0.0051        & 0.0002                                                              \\

12       & $\begin{pmatrix} 0.9648 & -0.0012 + 0.0014 i \\ -0.0012 -0.0014 i & 0.0612 \end{pmatrix}$  & $\begin{pmatrix} 0.9648 & 0.0612 \\ 0.0352 & 0.9388 \end{pmatrix}$ & $\begin{pmatrix} 0.9683 & 0.0651 \\ 0.0317 & 0.9349 \end{pmatrix}$ & 0.0074        & 0.0012                                                              \\

14       & $\begin{pmatrix} 0.9928 & 0.0004 + 0.0019 i \\ 0.0004 -0.0019 i & 0.0170 \end{pmatrix}$    & $\begin{pmatrix} 0.9928 & 0.0170 \\ 0.0072 & 0.9830 \end{pmatrix}$ & $\begin{pmatrix} 0.9915 & 0.0221 \\ 0.0085 & 0.9779 \end{pmatrix}$ & 0.0075        & 0.0004                                                              \\

23       & $\begin{pmatrix} 0.9713 & -0.0003 + 0.0049 i \\ -0.0003 -0.0049 i & 0.0407 \end{pmatrix}$  & $\begin{pmatrix} 0.9713 & 0.0407 \\ 0.0287 & 0.9593 \end{pmatrix}$ & $\begin{pmatrix} 0.9752 & 0.0450 \\ 0.0248 & 0.9550 \end{pmatrix}$ & 0.0082        & 0.0003                                                              \\
25       & $\begin{pmatrix} 0.9960 & 0.0038 + 0.0024 i \\ 0.0038 -0.0024 i & 0.0146 \end{pmatrix}$    & $\begin{pmatrix} 0.9960 & 0.0146 \\ 0.0040 & 0.9854 \end{pmatrix}$ & $\begin{pmatrix} 0.9961 & 0.0145 \\ 0.0039 & 0.9855 \end{pmatrix}$ & 0.0002        & 0.0038                                                              \\ \bottomrule
\end{tabular}%
}
\caption{Experimentally determined values in QDT. This includes estimates for the $|0\rangle\langle0|$ measurement, a comparison of calibration matrices constructed with both GST and prepare-and-measure methods, and an estimate of the quantumness of the measurement effect}
\label{tab:spam-table}
\end{table*}
The first three data columns respectively give direct estimates of the POVM 0-effect, the GST-derived calibration matrix, and the simplified calibration matrix. The final two columns respectively quantify from these the size of the approximation made from assuming perfect preparation and from assuming classical errors. The experiments were conducted with 8192 shots. Each of these values are small, both compared to the sampling error, and compared to the results in the main text.

We further examine these assumptions by conducting simulated experiments using the experimentally characterised data. Here, we aim to replicate the conditions of our results as much as possible in order to determine if there is any possibility of overestimating entanglement by using QREM. The procedure for this is as follows: 
\begin{itemize}
    \item Generate a set of possible 6-qubit lab states: $\rho = (1-p)\cdot \rho_{GHZ} + p\cdot \rho_{\text{noise}}$ for a range of $p$ and different noise (in particular: maximal mixtures and random states),
    \item Simulate out MQC with measurement error by using the GST estimates of the measurement probes $\{E_i\}$ to produce the complete coherence curve,
    \item Then, using the measure-and-prepare calibration matrix $A_{\text{local}}$, we apply QREM and invert the noise of the measurement effects,
    \item Finally, determine the final error-mitigated fidelity, and compare to the actual state $\rho$.
\end{itemize}
We performed this across a variety of different $\rho_{\text{noise}}$ and values for $p$, each sampled 500 times. The maximum average fidelity overestimate found with QREM was $6\times 10^{-4}$. We suspect that the MQC method of estimating state fidelity is particularly robust to erroneous QREM. The reason for this is that QREM carries the possibility of overestimating the number of $|0\rangle ^{\otimes n}$ readouts. However, MQC creates a curve measuring $|0\rangle^{\otimes n}$ amplitudes from peak to trough and then estimates the fidelity from the frequency here, which is more robust to translations of the entire curve.
\begin{figure*}
    \centering
    \includegraphics[width=\linewidth]{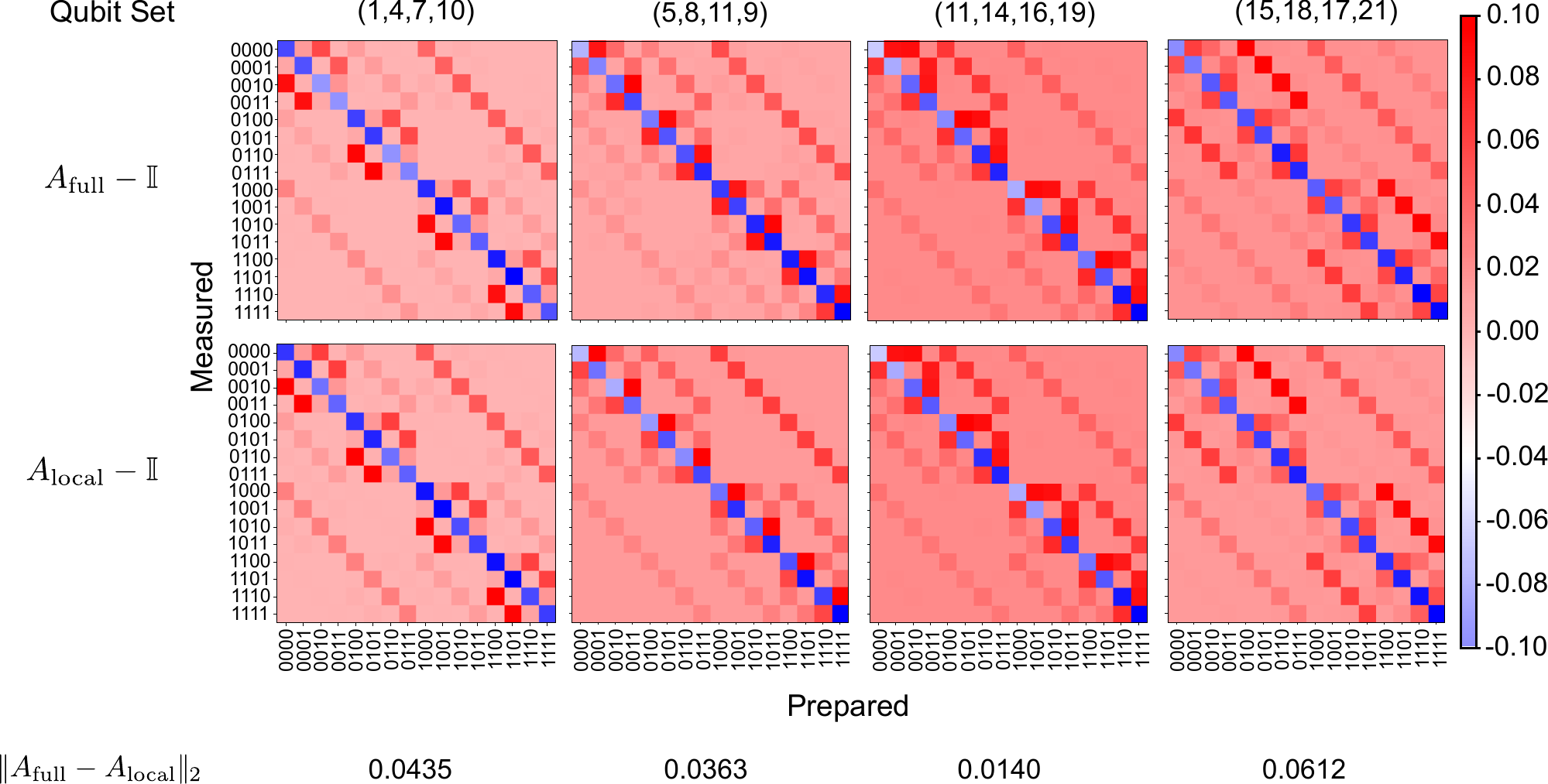}
    \caption{A comparison of measurement calibration matrices across samples of four over the device. We compare the full calibration matrix to that which is acquired by taking the tensor product of all local terms. In the matrix plots, we subtract off the identity matrix in order to better resolve finer details. }
    \label{fig:cal_matrix}
\end{figure*}
\subsection{Correlation of measurement error}
The final assumption made in the name of circuit reduction is that of locality of the errors. In the previous subsection, we justified that single-qubit measurement errors can be accurately characterised for the purpose of QREM on this hardware using only two circuits. An extrapolation to $n$ qubits would still require $2^n$ circuits to create the full calibration matrix. Instead, if we assume that measurement errors on different qubits are uncorrelated, we can take $A_{\text{local}} = \bigotimes_{i=1}^n A_i$ for the $n$ qubit system.\par
To check whether this is valid, we collected calibration matrices for four sets of four qubits and compared them to the tensor product of their marginals. Here, $A_{\text{full}}$ is the calibration matrix for all 16 prepared and measured states. Meanwhile, $A_{\text{local}}$ is the tensor product of local calibration matrices obtained from the preparation and measurement of $|0000\rangle$ and $|1111\rangle$. 
These results are summarised in Figure~\ref{fig:cal_matrix}. Here, we present matrix plots of each calibration matrix with the ideal case subtracted off. Each construction presents an almost identical depiction of the noise. One exception is the small occurrence of non-zero elements in $A_{\text{full}}$ for the set (15, 18, 17, 21) which is not present in $A_{\text{local}}$ -- however we believe this is not large enough to warrant accounting for. 
The difference between the local and full calibration matrices are quantified by their Frobenius distance. In all cases, this difference (representing the difference between 256 matrix elements) is small, and we are confident that results are not affected by making this simplifying assumption. 

\section{Using multiple quantum coherences to detect GME within noisy GHZ states\label{sec:MQC-explanation}}

GHZ states defined over arrays of qubits are more susceptible to noise than graph states defined over the same layout of qubits~\cite{briegel2001persistent}. So to detect entanglement, the graph state is generally the preferred option. However the GHZ state has a convenient structure that can be utilised for determining the more strict GME property, by calculating its fidelity. The fidelity between a desired pure state $\rho^\text{ideal}$ and the actual state $\rho$ is a similarity metric which can be written as
\begin{equation}
    F(\rho, \rho^\text{ideal}) = \text{Tr}(\rho \rho^\text{ideal}).
\end{equation}
When $\rho^\text{ideal}$ is a GHZ state, a fidelity value that is greater than or equal to 0.5 implies that $\rho$ is GME. The fidelity of a GHZ state can be written as 
\begin{equation}\label{eq:fidelity_p_plus_c}
    F = \frac{P + C}{2},
\end{equation}
where the \textit{population} $P = \rho_{00\ldots 0, 00\ldots 0} + \rho_{11\ldots 1, 11\ldots 1}$ can be directly measured as the sum of the \ket{00\ldots 0} and \ket{11\ldots 1} occupancies from the GHZ state and the \textit{coherence} $C = |\rho_{11\ldots 1, 00\ldots 0}| + |\rho_{00\ldots 0, 11\ldots 1}|$ can be indirectly measured through Multiple Quantum Coherences (MQC)~\cite{wei2020verifying} or parity oscillations~\cite{leibfried2005creation, monz201114}. In this work, we use MQC since a refocusing $\pi$-pulse similar to a Hahn echo~\cite{hahn1950spin} can be used to refocus low frequency noise and reduce dephasing~\cite{wei2020verifying}. This is because MQC is more robust to noise and scales better with respect to measurement error mitigation. MQC is a technique that has been adapted from solid state nuclear magnetic resonance~\cite{baum1985multiple} to many-body correlations and quantum information scrambling in trapped ions~\cite{wei2018exploring, garttner2017measuring}. More recently, it has been used to detect 18-qubit GME in the \textit{IBM Q System One} quantum device~\cite{wei2020verifying}. MQC works by utilising the amplified phase accumulated when phase rotating individual qubits. To help properly understand MQC and how it can be applied to calculate the fidelity, we will go through, in detail, steps to express the fidelity more concretely. The following working has been adapted from Wei \textit{et al}.~\cite{wei2020verifying} and G{\"a}rttner \textit{et al.}~\cite{garttner2017measuring}. 

We begin by writing down the following expression for the density matrix 
\begin{equation}
 \rho = \sum_{m, m^\prime=0}^N \rho_{m, m^\prime} |m\rangle \langle m^\prime |,   \label{eq:state-m-decomposition}
\end{equation}
where the basis states $\ket{m}$ are an equal superposition of each $N$-qubit state where there are $m$ qubits in the $|1\rangle$ state and $N-m$ in the $|0\rangle$ state. The state $\ket{m}$ satisfies
\begin{equation}
    \sum_{j=1}^N \frac{\sigma_z^j}{2} |m\rangle = A_m |m\rangle,\label{eq:m-eigenvalue}
\end{equation}
where the eigenvalues $A_m = m-N/2$ form the set $\{-N/2, -N/2+1, \ldots, N/2-1, N/2\}$. The density matrix can be expressed as a sum of different coherence sectors
\begin{equation}
    \rho = \sum_{q=-N}^N \rho_q,
\end{equation}
where
\begin{equation}\label{eq:rho_q}
    \rho_q := \sum_{m=0}^N \rho_{m, m-q} |m\rangle \langle m-q|,
\end{equation}
noting that $\rho_{i,j}=0$ when $i,j < 0$ or $i,j >N$. The coherence sectors $\rho_q$ account for the coherence between states $| m \rangle$ and $|m^\prime=m-q \rangle$ such that $A_m - A_{m^\prime} = q$. Note that $\rho$ being Hermitian implies that $\rho_q^\dagger = \rho_{-q}$, and each $\rho_q$ being orthogonal to one another implies $\text{Tr}(\rho_q \rho_p) = \delta_{q-p} \text{Tr}(\rho_{q}\rho_{-q})$. Although the states $\rho_q$ cannot be directly observed, we will show that the multiple quantum coherence amplitude defined as
\begin{equation}\label{eq:mqc-amplitude}
    I_q := \text{Tr}(\rho_q \rho_{-q}) = \text{Tr}(\rho_{-q} \rho_{q}) = I_{-q},
\end{equation} 
can. Since the density matrix of an ideal GHZ state only has non-zero values on each corner, the fidelity can be expanded as follows 
\begin{align}
    F &= \text{Tr}(\rho \rho^\text{GHZ}) \\
    &= \text{Tr}(\rho_{0} \rho_{0}^{\text{GHZ}}) + \text{Tr}(\rho_{N} \rho_{-N}^{\text{GHZ}}) + \text{Tr}(\rho_{-N} \rho_{N}^{\text{GHZ}}).\label{eq:fidelity-3-terms}
\end{align}
We will now work towards modifying the expression to only consist of directly measurable observables. By substituting values $0$ and $N$ for $q$ in Equation~\ref{eq:rho_q}, we get
\begin{align}
    &\rho_{0}^{\text{GHZ}} = \frac{1}{2}(|00 \ldots 0 \rangle \langle 00 \ldots 0 | + |11\ldots 1 \rangle \langle 11 \ldots 1|),\label{eq:rho_0^ghz}\\
    &\rho_{N}^\text{GHZ} = \frac{1}{2}|11 \ldots 1\rangle \langle 00 \ldots 0|,\label{eq:rho_minus_N}\\
    &\rho_{0} = \rho_{0,0}|00 \ldots 0 \rangle \langle 00 \ldots 0 | + \rho_{N,N}|11\ldots 1 \rangle \langle 11 \ldots 1|\\
    &\rho_{N} = \rho_{N,0}|11 \ldots 1\rangle \langle 00 \ldots 0|.\label{eq:rho_N}
\end{align}
From Equation~\ref{eq:rho_0^ghz}, due to orthogonality of the basis states, observe that
\begin{equation}
    \text{Tr}(\rho_0 \rho_0^\text{GHZ}) = \frac{1}{2}\left(\rho_{00\ldots 0,00\ldots 0} + \rho_{11\ldots 1, 11\ldots 1}\right),
\end{equation}
which is the half population term shown in Equation~\ref{eq:fidelity_p_plus_c}. The remaining two terms in Eq.~\ref{eq:fidelity-3-terms} will be expressed in terms of the amplitude $I_N$ introduced in Equation~\ref{eq:mqc-amplitude}. From Eqs.~\ref{eq:rho_minus_N} and \ref{eq:rho_N} we can write $\rho_N = \kappa \rho_N^\text{GHZ}$ for some complex constant $\kappa$. By substituting this into Eq.~\ref{eq:mqc-amplitude}, it can be shown that $\text{Tr}(\rho_N \rho_{-N}^{\text{GHZ}}) = \kappa I_N^\text{GHZ} = \frac{1}{\kappa^*} I_N$, where $I_N^\text{GHZ} = \text{Tr}(\rho_N^\text{GHZ} \rho_{-N}^{\text{GHZ}}) = \frac{1}{4}$. Thus $|\kappa| = 2\sqrt{I_N}$, hence $\rho_N = 2 \sqrt{I_N}\rho_N^\text{GHZ}$ since $\kappa$ can be made real by rotating $\rho$. Therefore, with similar working for $\text{Tr}(\rho_{-N} \rho_{N}^{\text{GHZ}})$ noting that $I_N = I_{-N}$ from Equation~\ref{eq:mqc-amplitude}, we have
\begin{align}
    &\text{Tr}(\rho_{N} \rho_{-N}^{\text{GHZ}}) + \text{Tr}(\rho_{-N} \rho_{N}^{\text{GHZ}}) \\
    &= 2\sqrt{I_{N}}(\text{Tr}(\rho_{N}^{\text{GHZ}} \rho_{-N}^{\text{GHZ}}) + \text{Tr}(\rho_{-N}^\text{GHZ} \rho_{N}^{\text{GHZ}}))\\
    &= \sqrt{I_N}.
\end{align}
Finally, we can rewrite the fidelity as
\begin{equation}
    F = \frac{1}{2}\left(\rho_{00\ldots 0,00\ldots 0} + \rho_{11\ldots 1, 11\ldots 1}\right) + \sqrt{I_N},
\end{equation}
with population $P = \rho_{00\ldots 0,00\ldots 0} + \rho_{11\ldots 1, 11\ldots 1}$ and coherence $C = 2\sqrt{I_N}$.

Now that we have expressed the coherence with respect to the quantum coherence amplitude $I_N$, we will show how $I_N$ can be measured. It can be calculated from the overlap signals $S_\phi := \text{Tr}(\rho_\phi \rho)$, which is a directly measurable quantity where $\rho_\phi := e^{-i\frac{\phi}{2}\sum_j \sigma_z^j} \rho  e^{i\frac{\phi}{2}\sum_j \sigma_z^j}$ is produced by rotating each qubit of the state $\rho$ by $\phi$ about the $Z$-basis. For an ideal GHZ state, this is equivalent to adding a phase of $N\phi$ to the state, that is $\frac{1}{\sqrt{2}}(|00\ldots 0\rangle + e^{-iN\phi} |11\ldots 1\rangle)$, which has the overlap signal reduce to
\begin{equation}
    S_\phi^{\text{ideal}} = \frac{1}{2}[1 + \cos (N\phi)],\label{eq:overlap-signal-ideal}
\end{equation}
where the constant term corresponds to the diagonal elements of the density matrix and the cosine term corresponds to the off-diagonal corner elements. To see the behaviour of the overlap signal when more states are included, we can consider the case of general pure states and express $\rho$ as
\begin{equation}
    \rho^\text{pure} := \sum_{m}^N a_m |m\rangle \sum_{m^\prime}^N \langle m^\prime | a_{m^\prime}^*, 
\end{equation}
where $a_m$ is the amplitude for the state $|m\rangle$ where $|m\rangle$ is defined as in Equation~\ref{eq:state-m-decomposition}. With the help of Equation~\ref{eq:m-eigenvalue}, we can write
\begin{align}
    \rho_\phi^\text{pure} &:= e^{-i\frac{\phi}{2}\sum_j \sigma_z^j} \sum_{m = 0}^N a_m |m\rangle \sum_{m^\prime = 0}^N \langle m^\prime | a_{m^\prime}^* e^{i\frac{\phi}{2}\sum_j \sigma_z^j} \\
    &= \sum_{m=0}^N a_m e^{-i\phi (m - N/2)} |m\rangle \sum_{m^\prime = 0}^N \langle m^\prime | a_{m^\prime}^* e^{i\phi (m^\prime - N/2)} \\
    &= \sum_{m=0}^N a_m e^{-i\phi m} |m\rangle \sum_{m^\prime = 0}^N \langle m^\prime | a_{m^\prime}^* e^{i\phi m^\prime}.
\end{align}
The overlap signal can then be reduced to the magnitude square of a complex Fourier series as follows
\begin{align}
    S_\phi^\text{pure} &= \text{Tr}[\rho_\phi^\text{pure} \rho^\text{pure}]\\
    &= \text{Tr}\left[\sum_{j=0}^N a_j e^{-i\phi j} |j\rangle \sum_{k=0}^N \langle k | a_{k}^* e^{i\phi k}    \sum_{l=0}^N a_l |l\rangle \sum_{m=0}^N \langle m | a_{m}^*\right] \\
    &= \sum_{j=0}^N |a_j|^2 e^{-i\phi j} \sum_{k=0}^N \langle k | a_{k}^* e^{i\phi k}    \sum_{l=0}^N a_l |l\rangle \\
    &= \left(\sum_{j=0}^N |a_j|^2 e^{-i\phi j}\right)  \left(\sum_{k=0}^N |a_{k}|^2 e^{i\phi k} \right)\\
    &=\left|\sum_{j=0}^N |a_j|^2 e^{-i\phi j}\right|^2.
\end{align}
When we substitute $a_0, a_N = 1/\sqrt{N}$ and $a_i = 0$ for $i\neq 0, N$, the expression reduces down to the ideal case for GHZ states shown in Equation~\ref{eq:overlap-signal-ideal}. By including amplitudes of non-all-zero and non-all-one states, lower frequencies are introduced into the behaviour and the $N$ frequency component is dampened.

The amplitude $I_N$ can be derived from $S_\phi$ as follows
\begin{align}
    S_\phi &= \text{Tr}\left( \sum_q e^{-i\frac{\phi}{2}\sum_j \sigma_z^j} \rho_q e^{i\frac{\phi}{2}\sum_j \sigma_z^j} \sum_p \rho_p \right)\\
    &= \text{Tr}\left( \sum_q e^{-i\phi((m - N/2) - (m - q - N/2))} \rho_q \sum_p \rho_p \right)\label{eq:use-m-eigenvalues}\\
    &= \sum_q e^{-i\phi q}\text{Tr}\left( \rho_q \sum_p \rho_p \right)\\
    &= \sum_q e^{-i\phi q}\text{Tr}\left( \rho_q \rho_{-q} \right)\\
    &= \sum_q e^{-i\phi q}I_q,
\end{align}
where Eqs. \ref{eq:m-eigenvalue} and \ref{eq:rho_q} are used to help obtain Equation~\ref{eq:use-m-eigenvalues}. Fourier transforming this result gives $I_q = \mathcal{N}^{-1}|\sum_\phi e^{i q \phi} S_\phi|$ where $\mathcal{N}$ is the number of angles $\phi$ within the summation. To ensure that the frequency $N$ is detectable, measure $S_\phi$ for $\phi = \frac{\pi j}{N+1}$, where $j=0, 1, \ldots, 2N + 1$ which can detect up to frequency $N+1$.
\newpage{}

\end{document}